# Structural, Electronic and Elastic Properties of zincblende III-Arsenide Binary Compounds: First-Principles Study

Umang Agarwal, Satish Chandra and Virendra Kumar
Department of Electronics Engineering
Indian Institute of Technology (Indian School of Mines)
Dhanbad 826 004, India
Email: umang.mathran@gmail.com

*Abstract*—First-principles calculations were performed, and the results from the study of structural, electronic and elastic properties of zincblende III-arsenide binary compounds (BAs, AlAs, GaAs and InAs) are presented. These properties have been calculated using an *ab initio* pseudopotential method based on density functional theory (DFT) with the local density approximation (LDA) for the exchange-correlation potential. The results obtained for the calculated properties have been compared with experimental data and other computational works. It has also been found that our results with LDA are in good agreement with other computational work wherever these are available.

*Keywords—first-principles; local density approximation; density functional theory; III-V compounds; structural properties; electronic properties; elastic properties*

## I. INTRODUCTION

Over the past few decades, Group III-V have been extensively studied. The Group III-Arsenide compounds of this family have applications in electronic and optoelectronic devices like light emitting diodes (LEDs), photo detectors, lasers, modulators, integrated circuits and filters [1]. Under normal conditions, these compounds crystallize in the zincblende (ZB) structure [2]. Out of the four compounds, Boron Arsenide (BAs) and Aluminum Arsenide (AlAs) are indirect band gap with values of 1.46 eV and 2.24 eV respectively, whereas Gallium Arsenide (GaAs) and Indium Arsenide (InAs) are direct band gap semiconductors with band gaps of 1.424 eV and 0.42 eV respectively [3, 4, 5]. Recently, the authors have studied the ZB-AlN semiconductor by using first-principles calculations [6]. In the present work, we have investigated the structural paramters, electronic and elastic properties of BAs, AlAs, GaAs and InAs crystals in zinc-blende structure at room temperature (300K) and normal pressure. First-principles calculations were performed within the density functional theory (DFT) framework [7, 8] using an open-source software suite, Quantum ESPRESSO [9], with local density approximation (LDA) for the exchange-correlation functional. The results accomplished are correlated with the existing theories and experiments. A fairly good agreement has been made between them.

## II. COMPUTATIONAL METHODS

The atoms in zincblende structure are in face-centered cubic (FCC) positions as R (0, 0, 0) and X (1/4, 1/4, 1/4), where R ≡ B, Al, Ga, In and X ≡ As. The interactions between ions and valence electrons have been described using the norm-conserving pseudopotentials (NCP) method. The exchange-correlation effects were treated by the local density approximation (LDA) given by Perdew-Zunger (PZ) [10] using the schemes of Von Barth-Car for for B, Al, Ga, In and As. The k-meshes were sampled according to a Monkhorst-Pack [11] scheme with spacing of 0.5/Å as follows: 9×9×9 for computing structural properties, energy band gap and elastic properties, whereas 32×32×32 for generating the density of states.

The electronic configurations for the B, Al, Ga, In and As are [He]$2s^2 2p^1$, [Ne]$3s^2 3p^1$, [Ar]$3d^{10} 4s^2 4p^1$, [Kr]$4d^{10} 5s^2 5p^1$ and [Ar]$3d^{10} 4s^2 4p^3$, respectively. The kinetic energy cut-offs for wave-functions were taken as 135 Ry, 95 Ry, 95 Ry and 100 Ry for BAs, AlAs, GaAs and InAs, respectively. The kinetic energy cut-offs for charge-densities were taken to be 540 Ry for BAs, 380 Ry for AlAs and GaAs and 400 Ry for InAs. The self-consistent calculations converged when total energy of the system was stable within $< 10^{-8}$ Ry.

The elastic stiffness coefficients ($C_{ij}$) and bulk ($B$), shear ($G$) and Young's ($E$) moduli and Poisson ratio ($\gamma$) for the binary compounds were calculated using a tool ElaStic [12] along with Quantum ESPRESSO code. The stress approach was implemented which defines the lattice deformation types according to [13].

## III. RESULTS AND DISCUSSION

### A. Structural Parameters

The equilibrium structural parameters of the compounds were investigated by calculating and fitting the total energy for a unit cell as a function of volume to the Murnaghan's equation of state [14]:

$$E(V) = E_0 + \frac{B_0 V}{B_0'}\left(\frac{(V_0/V)^{B_0'}}{B_0'-1} + 1\right) - \frac{B_0 V_0}{B_0'-1} \qquad (1)$$

where $E_0$ is the equilibrium energy, $B_0$ is the bulk modulus, $B'_0$ is the first derivate of $B_0$. The total energy vs unit cell volume of BAs, AlAs, GaAs and InAs are shown in Fig. 1 to 4. The lattice constant $a$, bulk modulus $B_0$ and its pressure derivative $B'_0$ were obtained from energy-volume curves and are listed in Table I along with the available experimental and reported data. Comparing our results with the available published works, a good agreement was established. It was noted that the computed lattice constants using LDA is lower than the experimental value by 0.91% for BAs, 0.86% for AlAs, 1.89% for GaAs and 3.14% for InAs. This is in agreement with the general trend that LDA underestimates the lattice constants. The bulk modulus $B_0$ and its pressure derivate $B'_0$ are also shown in the Table I and are found to be in good agreement with the experimental and previously reported results.

### B. Electronic Properties

Although the computed LDA lattice constants are close to the corresponding experimental lattice constants but due to the high sensitivity of electronic band gap to the lattice parameter, the experimental lattice parameter was chosen for the consistent study of electronic structure using the exchange-correlation approximation. The band structure, energy band gap and density of states (DOS) have been calculated.

The obtained electronic band structures and density of states (DOS) are presented in Fig. 5 to 12. For all the compounds, the maxima of the valence band occurs at $\Gamma$ point. The minima of the conduction band occurs at different positions for the compounds. For BAs, the minima occurs at a point between $\Gamma$ and $X$ which we call $\delta$, while for AlAs, it appears at a point in $X$. Hence, BAs and AlAs have indirect band gap along $\Gamma - \delta$ and $\Gamma - X$, respectively. The minima of the conduction band occurs on $\Gamma$ for both GaAs and InAs; and hence both have direct band gap along $\Gamma - \Gamma$.

Using the obtained band structures (Fig. 5 to 8), the corresponding band gaps were found and are listed in Table I along with the available experimental data and reported results. A known effect of LDA is the underestimation of the band gap which was observed. However, our results were in accordance to various other previously reported band gaps using LDA.

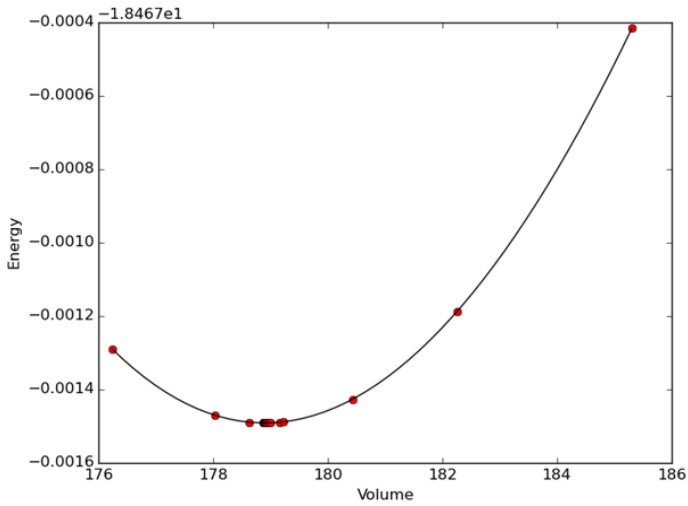

Fig. 1. Energy (in Ry) vs volume (in a.u.³) graph of BAs

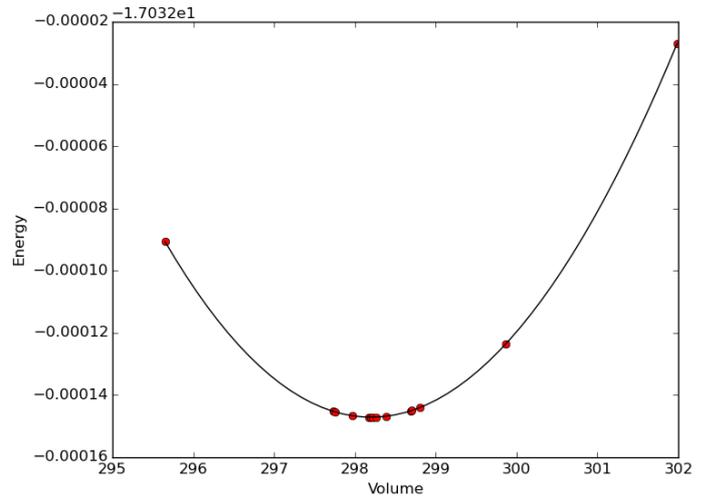

Fig. 2. Energy (in Ry) vs volume (in a.u.³) graph of AlAs

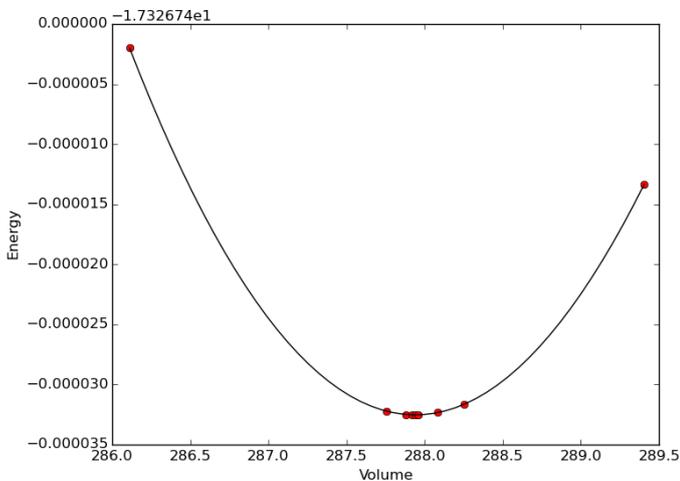

Fig. 3. Energy (in Ry) vs volume (in a.u.³) graph of GaAs

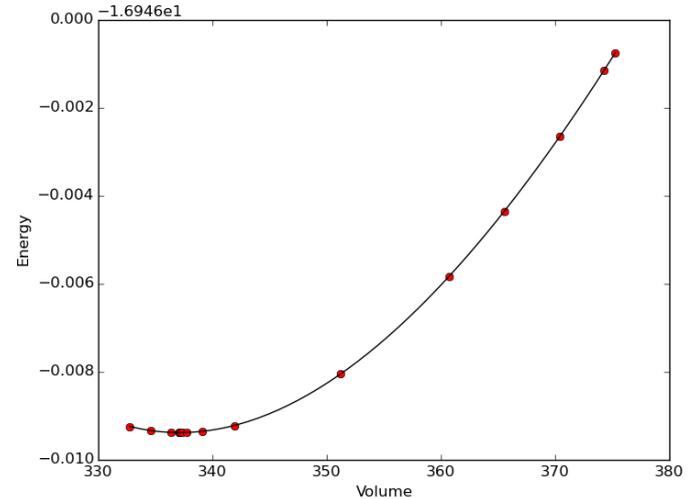

Fig. 4. Energy (in Ry) vs volume (in a.u.³) graph of InAs

TABLE I. LATTICE PARAMETER A (IN Å), BULK MODULUS B (IN GPa$^3$), PRESSURE DERIVATIVE OF BULK MODULUS B′ AND ENERGY BAND GAP E$_G$ (IN EV) OF BAs, AlAs, GaAs AND InAs

| Compound | Lattice Parameter | | | Bulk Modulus | | | Derivative of Bulk Modulus | | | Energy Band Gap | | |
|---|---|---|---|---|---|---|---|---|---|---|---|---|
| | *This work* | *Expt. value* | *Theoretical values* | *This work* | *Expt. value* | *Theoretical values* | *This work* | *Expt. values* | *Theoretical values* | *This work* | *Expt. value* | *Theoretical values* |
| BAs | 4.7332 | 4.777[a] | 4.743[c], 4.812[c], 4.741[d] | 146.26 | - | 152[c], 147.5[d] | 4.111 | 3.984[g] | 4.216[d] | 1.1160 ($\Gamma - \delta$) | 1.46[h] | 1.21[c], 1.13[d] |
| AlAs | 5.6121 | 5.661[b] | 5.633[d], 5.614[e], 5.644[f] | 75.11 | 82[g] | 75.1[d], 74.7[e] | 4.138 | 4.182[g] | 4.512[d] | 1.3509 ($\Gamma - X$) | 2.24[h] | 1.31[d] |
| GaAs | 5.5469 | 5.654[a] | 5.608[d], 5.530[e], 5.649[f] | 76.40 | 77[g] | 75.2[d], 75.7[e] | 4.295 | 4.487[g] | 4.814[d] | 0.8836 ($\Gamma - \Gamma$) | 1.424[h] | 0.32[c] |
| InAs | 5.8463 | 6.036[a] | 6.030[d], 5.921[e], 6.015[f] | 71.85 | 58[g] | 60.9[d], 61.7[d] | 4.719 | 4.79[g] | 4.691[d] | 0.2372 ($\Gamma - \Gamma$) | 0.42[h] | -0.64[f], 0.40[d] |

[a.] Reference [15]
[b.] Reference [16]
[c.] Reference [17]
[d.] Reference [1]
[e.] Reference [18]
[f.] Reference [19]
[g.] Reference [20]
[h.] References [3], [4] and [5]

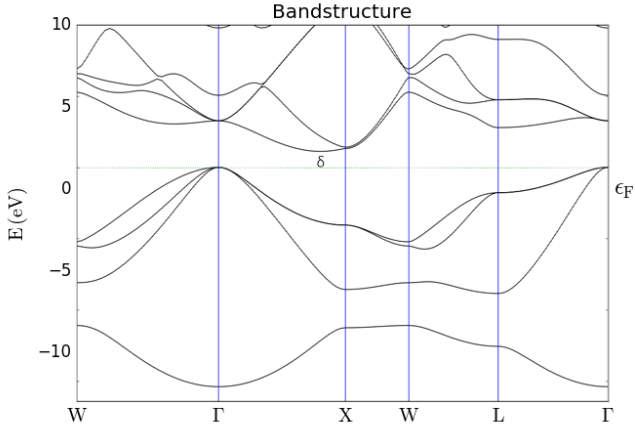

Fig. 5. Band structure of BAs. The energy band gap is along ($\Gamma - \delta$).

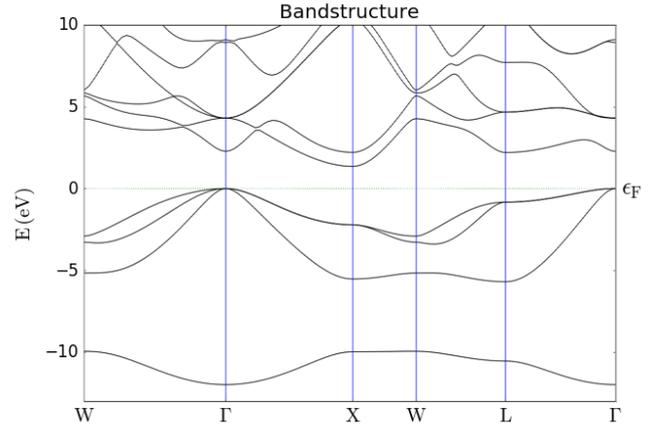

Fig. 6. Band structure of AlAs. The energy band gap is along ($\Gamma - X$).

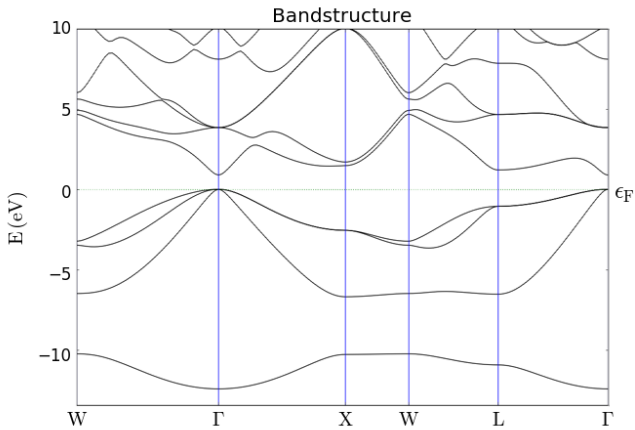

Fig. 7. Band structure of GaAs. The energy band gap is along ($\Gamma - \Gamma$).

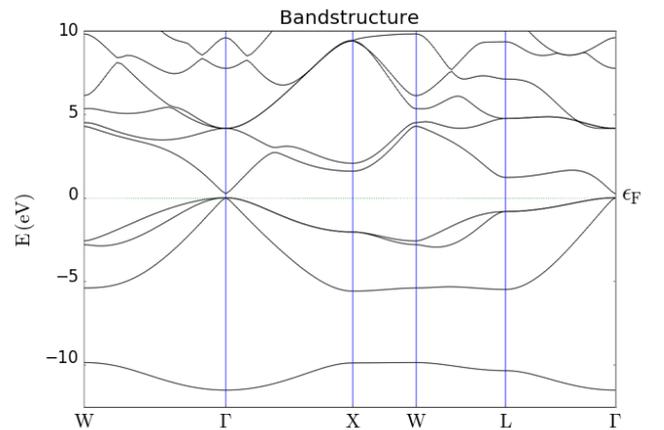

Fig. 8. Band structure of InAs. The energy band gap is along ($\Gamma - \Gamma$).

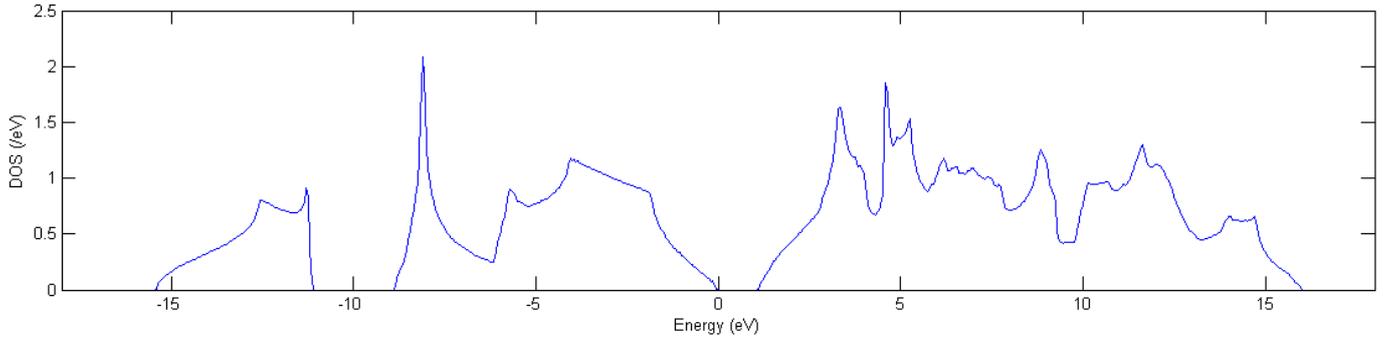

Fig. 9. Total density of states (DOS) for BAs

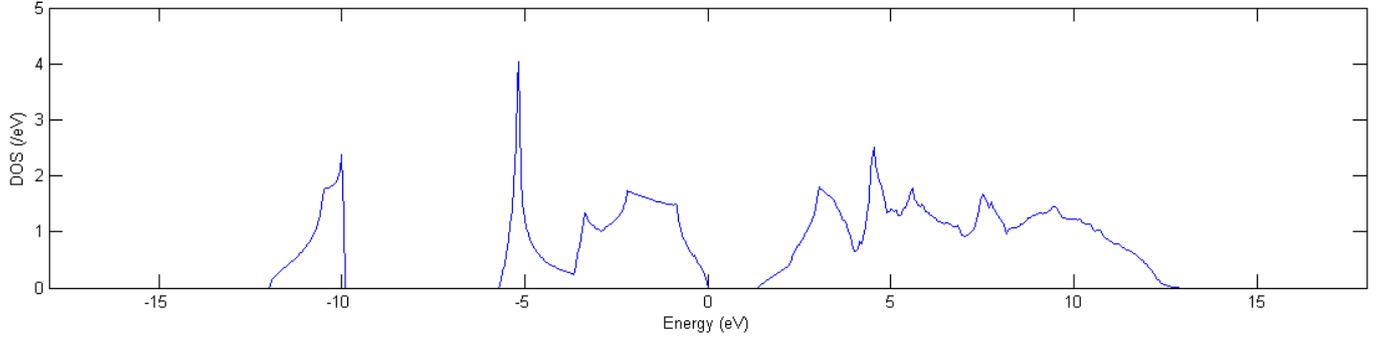

Fig. 10. Total density of states (DOS) for AlAs

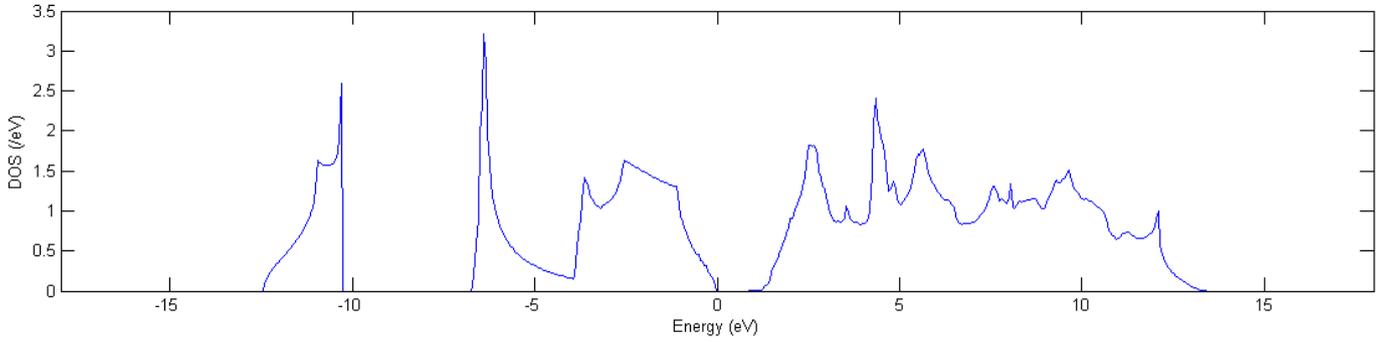

Fig. 11. Total density of states (DOS) for GaAs

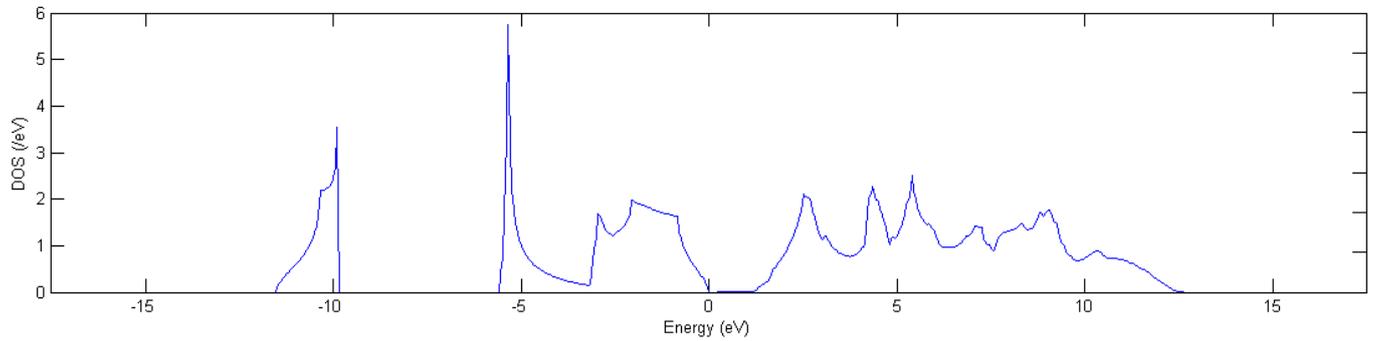

Fig. 12. Total density of states (DOS) for InAs

## C. Elastic Properties

The elastic stiffness constants ($C_{11}$, $C_{12}$ and $C_{44}$) for the zincblende III-As compounds are listed in Table II along with the experimental and previously calculated data. For a cubic crystal system, the mechanical stability under isotropic pressure can be judged from the following conditions [21]:

$$C_{11} - C_{12} > 0, \ C_{11} + 2C_{12} > 0, \ C_{44} > 0 \qquad (2)$$

It was found that the elastic stiffness constants ($C_{ij}$) of the compounds were consistent with (2) indicating that the zincblende structures of these compounds are stable. Using the Voigt-Reuss-Hill averaging scheme with the calculated results of single crystal elastic constants, the bulk modulus ($B$), shear modulus ($G$), and Young's modulus ($E$) were determined and are listed in Table II along with previous reported values.

TABLE II. ELASTIC STIFFNESS CONSTANTS $C_{ij}$ (IN GPA), BULK MODULUS B (IN GPA), SHEAR MODULUS G (IN GPA), YOUNG'S MODULUS E (IN GPA) AND POISSON RATIO OF BAs, AlAs, GaAs AND InAs

| Compound | Parameter | This work | Reported values |
|---|---|---|---|
| BAs | $C_{11}$ | 291.4 | 301.26[i], 295[i] |
| | $C_{12}$ | 73.7 | 77.23[i], 78[i] |
| | $C_{44}$ | 166.6 | 163.87[i], 177[i] |
| | $B$ | 146.24 | 151.91[i], 152[i] |
| | $G$ | 140.48 | - |
| | $E$ | 319.22 | - |
| | $\gamma$ | 0.14 | - |
| AlAs | $C_{11}$ | 113.2 | 119.9[j], 116[k] |
| | $C_{12}$ | 56.1 | 57.5[j], 55[k] |
| | $C_{44}$ | 59.1 | 56.6[j] |
| | $B$ | 75.18 | 75[k] |
| | $G$ | 44.16 | 31[k] |
| | $E$ | 110.78 | - |
| | $\gamma$ | 0.25 | 0.324[j] |
| GaAs | $C_{11}$ | 123.8 | 138[k], 106.5[l] |
| | $C_{12}$ | 51.8 | 55[k], 60.2[l] |
| | $C_{44}$ | 71.3 | 66[k], 33.6[l] |
| | $B$ | 75.80 | 75.6[l] |
| | $G$ | 54.19 | 28.9[l] |
| | $E$ | 131.29 | 63.0[l] |
| | $\gamma$ | 0.21 | 0.36[l] |
| InAs | $C_{11}$ | 71.3 | 83.4[m] |
| | $C_{12}$ | 34.9 | 45.4[m] |
| | $C_{44}$ | 41 | 39.5[m] |
| | $B$ | 47.01 | 58.1[m] |
| | $G$ | 29.61 | 19.0[m] |
| | $E$ | 73.42 | 51.4[m] |
| | $\gamma$ | 0.24 | 0.35[m] |

[i.] References [15] and [22]
[j.] Reference [23]
[k.] Reference [24]
[l.] Reference [25]
[m.] Reference [26]

## IV. CONCLUSION

We have performed first-principles calculations for zincblende Group III-Arsenide compounds and determined their equilibrium structural parameters, electronic and elastic properties. The calculated band gap values are lesser than the experimental values due to the intrinsic property of LDA pseudopotentials. This is due to a simpler form of the exchange-correlation functional. The elastic stiffness constants ($C_{11}$, $C_{12}$ and $C_{44}$), and elastic moduli of the semiconductors have been calculated and found to be in good agreement with the reported values. Thus, our calculated values are in fair agreement with the available reported values, which shows the usefulness of the open-source software suite, Quantum ESPRESSO.